\documentclass[%
 reprint,
 amsmath,amssymb,
 aps,
]{revtex4-2}

\usepackage{graphicx}
\usepackage{dcolumn}
\usepackage{bm}
\usepackage{xcolor}
\usepackage{float}

\newcommand{\quotes}[1]{``#1''}
\begin{document}

\preprint{APS/123-QED}

\title{A Journey in Implementing Computational Physics from the Ground Up
}


\author{Maria C. Babiuc Hamilton}
\email{babiuc@marshall.edu}
\affiliation{
Marshall University, Department of Mathematics and Physics, \\
 Huntington, WV, United States
}

\date{\today}

\begin{abstract}
This chapter narrates the journey of developing and integrating computing into the physics curriculum through three consecutive courses, each tailored to the learners' level. It starts with the entry-level \quotes{Physics Playground in Python} for high school and freshman students with no programming experience, designed in the spirit of the "Hello World" approach.
At the sophomore and junior level, students from all sciences and engineering disciplines learn \quotes{Scientific Computing with Python} in an environment based on the “Two Bites at Every Apple” approach.
Ultimately, upper undergraduate and entry-level graduate students take \quotes{Computational Physics,} to develop their skills in solving advanced problems using complex numerical algorithms and computational tools.
This journey showcases the increasing complexity and sophistication of computational tools and techniques that can be incorporated into the physical science curriculum, serving as a guide for educators looking to integrate computing into their teaching. It also aims to inspire students by showcasing the impact and potential of computational methods in science education and research.
\end{abstract}

\maketitle

\section{\label{sec:intro}Introduction}

\paragraph{Importance of Computational Physics in Modern Education}
Computational physics, which uses computational methods and algorithms to solve complex physical problems, has become an indispensable part of the field, alongside theory and experiment. This approach enables and supports new discoveries, such as simulating the dynamics of galaxies, predicting the behavior of complex fluids, or modeling climate changes. Computational physics allows scientists to explore scenarios and test theories at scales and speeds not feasible in traditional laboratory settings. It has supported significant breakthroughs in physics, such as the first complete simulation of black hole merger \cite{Pretorius2005} in my research area.

Incorporating computation into the physical sciences curriculum is essential to keep pace with current trends in research, industry, and society. However, formal programs training students in computational physics are absent from most departments \cite{Phillips2023, Caballero2024}. While physics faculty recognize the importance of an educational experience that includes computation, the vast majority do not include it in their courses. Furthermore, the recent steep decline in the number of physics majors highlights the urgent need for a strategic plan to update the curriculum, allowing for future growth in both students and faculty and reflecting the job market realities for physics majors.

It is well known that science and engineering students often avoid computation classes, while computer science students lack the necessary mathematics and science background to simulate natural processes \cite{Hamerski2022}. Adding a succession of computation courses in the Physics curricula could benefit all Science, Engineering, and Mathematics majors, filling a crucial gap in their preparation for the job market, which relies heavily on computational skills \cite{Behringer2017, ACM2024}. The increase in computer modeling use in science and industry underscores this need, with applications ranging from flight simulation to weather forecasting and disease spread prognostics.
Computational skills, important for thriving in the real world, are especially crucial now with the rise of data science, machine learning, and artificial intelligence. These technologies are transforming industries and research fields, making computational proficiency a valuable asset for students in physics and beyond.

\paragraph{Vision for integrating computational physics}
Physics education traditionally relies on lectures, laboratory experiments, and exercises because they effectively teach fundamental principles, concepts, and problem-solving skills. This approach builds a solid foundation but offers limited interaction and real-time feedback, especially in exploring phenomena like cosmic events, quantum mechanics, or high-energy physics.

Computational physics is not meant to replace traditional learning but to enhance it. By simulating complex systems, testing and visualizing abstract concepts, analyzing experimental data, and planning future experiments, computational physics broadens our understanding of physical phenomena.

In the evolving landscape of science and technology, computational physics ensures that education equips students not only with physical knowledge but also with the ability of applying computational thinking to real world problems. We must meet students where they are and guide them to reach their educational goals and potential, preparing them for job market demands and scientific frontiers.

Scaffolded learning, an instructional method that provides successive support steps, helps students develop skills progressively and achieve higher understanding levels. As students advance, supports are gradually removed to encourage independent learning. This approach promotes equity by making the curriculum accessible to all students, ensuring a strong foundational understanding before advancing. Students build confidence by mastering simpler tasks before moving on to more complex ones, reducing frustration and enhancing comprehension \cite{Caballero2012, Caballero2012MI}.

After exploring the motivation behind integrating computational physics into the physical science curriculum, the reader should be persuaded of the importance and impact of computational physics in modern physics education. 
Let us now take a closer look at the challenges and opportunities of integrating computing into the physics curriculum, and present the scaffolded approach we adopted, describing, step by step, the ladder we built to help students acquire the skills needed for using and developing programs to study physical phenomena. My aim is to support various levels of complexity, catering to both beginners and advanced learners, and to encourage active participation through live coding, immediate feedback, and interactive visualizations.

\paragraph{Personal Motivation}
My home college is Marshall University, a regional, state-supported, mid-sized R2 research university located in Huntington, WV. The percentage of West Virginians who earn a college degree is only 12.4\%. Moreover, the local students in the science technology, engineering, and mathematics (STEM) disciplines constitutes a vastly underrepresented group among our students, who have a predilection to major in Nursing or Psychology.
I describe below the stages implemented in the curriculum development at my university, that have the potential to lead to an increased number of students in the STEM fields.  

Similarly to many colleagues who teach computation into physics courses \cite{Young2019}, I was not formally trained in computation.
Instead, I learned scientific software and programming by incorporating computer modeling into my research in computational astrophysics, simulating black hole and neutron star collisions, and calculating the energy released in gravitational waves \cite{Babiuc2004, Babiuc2005, Motamed2006}.
My first intentional exploration of this curricular change was prompted by a discussion with a colleague who was interested in my experience involving undergraduate students in computational physics through my research \cite{Buskirk2018}. 
At that time, I had already introduced a junior/senior-level laboratory-based course in \quotes{Computational Physics} for students interested in using programming languages to simulate physical processes. I was using Mathematica and C++ in that course but started experimenting with Python because it is easier to learn and debug. However, I was unaware of the Jupyter notebook environment, which is an ideal platform for introducing students to computational physics \cite{Odden2019, Odden2020}.

I started developing material using this platform at the height of the COVID pandemic, when all teaching moved online and there was an acute need to enhance the virtual learning environment. I received an invitation to be part of the Summer 2020 Immersion Team for first-generation freshman students, where I designed and presented a set of virtual physics experiments during a two-week program. I used a combination of PhET Simulations and Jupyter notebooks, allowing students from various STEM disciplines to explore and interact with physical phenomena such as gravitational attraction, wave motion, gas dynamics, and electric interactions. 
This is when I became a member of the Integration of Computation into Undergraduate Physics (PICUP) community \cite{PICUP2024} and had the opportunity to attend a PICUP Virtual Workshop. . During the workshop, I received invaluable help in redesigning my Computational Physics course around a more accessible, project-based curriculum. This experience planted the seeds for a sequential offering of three computational physics classes, from freshman to senior/master's students, forming the backbone of a new minor in Computational Physics.

The excellent resources available on the PICUP webpage provided an ideal entry point for me, along with the wealth of information online about scientific computing and computational physics courses offered by other universities \cite{PERL2024, CCSE2024}. The number of useful web references is vast, and space does not permit listing all the excellent resources. 
One of the most remarkable curricula to emulate is the one at Brigham Young University \cite{Computational2024}
and the article \quotes{Teaching computational physics as a laboratory sequence} by Ross L. Spencer describing it \cite{Spencer2005}.
Other excellent resources are provided by the Teaching to Increase Diversity and Equity in STEM (TIDES) initiative at the Bryn Mawr College \cite{BrynMawr2024}, as well as the textbooks {\em Computational Modeling and Visualization of Physical Systems with Python} by Jay Wang \cite{Wang2016}, and {\em Basic Concepts in Computational Physics} by B. A. Stickler and E. Schachinger \cite{Stickler2014} along with the open-source programs that come with these books.

The guide in \quotes{Effective Practices for Physics Programs} offers valuable steps for improving the physics curriculum, and how to intentionally integrate computation into the physics program \cite{Irving2017, Irving2020}.
Useful information that describes curriculum changes necessary to introduce computational physics courses are also found in the works of Christopher J. Burke, and Timothy J. Atherton \cite{Burke2017} who developed a project-based computational physics course focused on problem solving. 

\section{\label{sec:implement}Implementation Stages}

Computational tools have learning curves that can differ significantly among students. Developing proficiency takes time and can be both a blessing and a curse: it can make science learning more engaging and enjoyable but can also lead to frustration and aversion for some.
Instructors often face a learning curve as well, frequently having to learn numerical analysis and computer programming on the fly while teaching their students \cite{Gambrell2024}.

To account for these factors when designing the curriculum, I drew inspiration from the excellent open-source handbook {\em Teaching and Learning with Jupyter} \cite{Jupyter2023}, 
which provides valuable insights and practical strategies for integrating computational tools into the curriculum, based on 
 a scaffolded, innovative approach, ensuring flexibility and adaptability to a breadth of skills.

Let's start with detailing the implementation of the \quotes{Physics Playground in Python} course targeted to high school seniors and college freshmen, who can explore simple physics concepts using Python, in an incremental, hands-on scaffolded learning \quotes{Hello World} approach. 
This assumes no previous knowledge of computation or physics, and first, students are introduced to the Python programming language and learn the basic physics principles of motions, through interactive Jupyter notebooks. They progress from learning simple programming rules and syntax to implementing simulations of bouncing balls, adding complexity step by step, culminating in implementing complex systems such as planetary orbits.

This is followed by a \quotes{Scientific Computing in Python} course taken later, by sophomores and juniors, where students are exposed to a range of computational tools and languages, assuming limited familiarity with programming. Here, students learn about powerful Python libraries such as NumPy, SciPy, and Matplotlib, and gain insights into developing a broader scientific computing language, integrating mathematical models with real-world problems. The approach used is \quotes{Two Bites at Every Apple,} revisiting and reinforcing concepts for deeper understanding and promoting active learning.

We end with the \quotes{Computational Physics} course designed for upper-level undergraduates and graduate students, that follows the \quotes{Win-day-one} approach to engage students from the start with complex problems and immediately immerse them in high-level computational work and numerical algorithms. This allows students to explore complex simulations, such as the restricted three body problem and independent research projects such as black hole mergers and gravitational waves.

As a strand linking this narrative, I present a selection of students' accomplishments along this journey of learning computational physics. These achievements illustrate the progression from basic to advanced levels across the three courses, where computational skills are built and applied in increasingly complex scenarios, ranging from simulating a simple bouncing ball to modeling black hole and neutron star collisions.

The conclusion summarizes our approach and explores possible future directions for integrating computational physics into education. 
This includes discussing the potential impact of the increasing use of Large Language Models (LLMs) on teaching computational physics and their impact on student preparation. 

\subsection{\label{sec:ppp}Physics Playground in Python}
\subsubsection{Setting the stage}

The journey starts with an introductory curriculum for students who had no previous exposure to physics or to computational programming languages. 
\quotes{Physics Playground in Python} is aimed at students who take their first steps into computational physics and implements the \quotes{Hello World} approach, by engaging them in hands-on activities and interactive coding, with checkpoints to solidify understanding and build confidence as they gradually increase in complexity. 
Students start with simple simulations, such as a bouncing ball, and gradually progress to more complex systems like planetary orbits. This course lays the groundwork by offering an intuitive understanding of physics through computational experimentation, using tools like Jupyter notebook and GitHub to foster an interactive and collaborative learning environment. 
To enhance their engagement, students use VPython, a Python library that makes it easy to create 3D animations and visualizations of physical systems \cite{VPython2024}.

The objective is to build confidence in students' ability to write their first computer program, have fun, and step into the world of computing with minimal background knowledge.

The curriculum is divided into two modules: physics and computation. For the physics part, students are introduced to motion through a series of short, interactive mini-lectures, starting with motion at constant velocity in one dimension and adding acceleration. This is followed by exploring constant motion in two and three dimensions, and concluded with motion under variable acceleration, beginning with examples like cannonballs and satellite launches, and culminating in complex systems such as the solar system.

In the computational module, students begin by learning basic syntax, variable assignment, arithmetic operations, loops, conditional statements, and functions. They run and debug simple scripts written in Python and VPython, using Jupyter notebooks within the Anaconda software \cite{Anaconda}. This setup provides a friendly, readable implementation of code, interwoven with documentation and simple exercises.

Assessment is based on a portfolio rather than grades, allowing students to take pride in building their programs from simple to complex and showcasing the results. This approach mirrors real-world job settings.

To facilitate access to course materials and enhance collaboration, I created a GitHub repository to host the modules \cite{GitHub}. Using GitHub Classroom, available through GitHub Education, I set up a dedicated classroom that allows students direct access to the material online at \href{https://github.com/mbabiuc/computationalphysicsmu}{github.com/mbabiuc/computationalphysicsmu}.
This platform provides a centralized location for all course content, including code examples, exercises, and documentation, and serves as a repository for students' portfolios, enabling them to track their progress and showcase their work. This setup fosters an interactive and collaborative learning environment, where students can easily clone the repository to their local machines, submit assignments, and collaborate with peers. Through this process, they learn essential skills in version control and collaborative coding practices.

\quotes{Physics Playground in Python} is offered online as a summer internship for high school students interested in learning more about physics and space, using the Python programming language within the Jupyter notebook interactive environment. No prior coding experience is required, but a strong background in mathematics is highly recommended.

The schedule is highly flexible, consisting of a 4-hour block per day for one to four weeks, according to the students' convenience. The only requirement is to find at least one overlapping hour per day among all students to present their work. 

This hands-on approach can easily be restructured as a one-semester, two-credit-hour, project-based introductory course to computer simulations in physics for freshman and sophomore students.

\subsubsection{First Steps}

One week before the course begins, students receive an email from the instructor containing a welcome message and a brief biography. The email outlines what to expect from the course and includes a link to an online scheduling tool, such as Doodle, to arrange a mutually convenient time for meetings. Additionally, the email provides detailed instructions on how to install Anaconda and VPython, ensuring that students are prepared to start with the necessary software.

The first meeting is reserved for student introductions and troubleshooting the installation of Anaconda on their computers. Anaconda includes Jupyter, and it is easy to install on Windows, Mac, and Linux. 
Once installed, students can jump right into learning Python and VPython in an interactive environment. This process usually goes smoothly, with the main issue being if students do not have installation rights on their laptops. As a workaround, students can use interactive online Jupyter platforms such as Google Colab, JupyterHub, or MyBinder.
By the end of the first meeting, students are set up with the necessary tools and ready to embark on their computational physics journey.
The rest of the first day is dedicated to helping students take their first steps in becoming familiar with the Python programming language by working through the initial TIDES modules. They are designed not only to introduce students to Python but also to engage them in a way that makes learning enjoyable and relatable. 
Each module starts with a joke or a profile of a scientist and includes regular \quotes{breakpoints} for reflection and individual exercises. 
The interactive nature of the modules, combined with the reflection points, ensures that students can understand and apply the concepts effectively while building confidence in their programming skills.
Students are encouraged to build their own Jupyter notebooks to answer the questions and complete the exercises, marking the first addition to their portfolio.

The first module, Module 00, is available only in PDF format and introduces students to computing while offering various programming resources. Module 01 is a self-contained introduction to programming in Python and is divided into three parts: 01A, 01B, and 01C. Each part takes roughly three hours to complete, occupying most of the remaining time during the first week.

Each day, except for the first, the instructor dedicates 10-15 minutes of the hour when all students are present to introduce the physics of motion. 
The journey begins with a brief history of motion, starting with Aristotle—the first physicist—who around 400 BC instigated the misconception that objects of different masses fall at different rates, a belief still held by many today.
It took almost two millenia for Galileo to conduct his famous experiment around 1590, demonstrating that the time of descent was independent of mass by dropping unequal weights of the same material from the Leaning Tower of Pisa.
I then establish the mathematical basis of motion by introducing formulas for one-dimensional motion with constant velocity and constant acceleration.

The remainder of this time is used to answer students' questions. The last day of the first week is dedicated to an introduction to VPython, which students will use to create simulations that illustrate motion, forces, and vectors, helping them understand these phenomena through visual representation.

To avoid overwhelming the students, they are provided with a minimally-working VPython program that compiles without errors and creates a basic visualization. Students learn to create their first canvas and design objects such as sphere and walls, as well as how to animate these objects. All students achieve this milestone, and those registered for only one week leave with the satisfaction of having accomplished their first simulation of a ball bouncing off two walls.

\subsubsection{Simple Physics Simulations}

The second week begins with an introduction to scientific computation and code development. Students work through Module 03, which covers simple iterative numerical methods to update the variables, and implement the two-dimensional motion, starting with constant velocity, then progress by adding constant acceleration. This module takes about six hours to complete. By the end of it, students apply their newly acquired knowledge to their VPython programs. They first simulate a ball in 2D bouncing off the walls of a 3D box, and continue with simulating a projectile motion. 
 
During the physics mini-lectures, students learn about Newton, who laid the foundations of mechanics with his second law of dynamics and understand the concept of force. 
This knowledge is applied to explain the effects of drag forces on moving objects, allowing students to incorporate air resistance in their projectile simulations and build more realistic models.

Students who choose to conclude the internship at the end of the second week will have a solid understanding of simulating simple motion.
Their portfolio will include the bouncing ball in one and two dimensions, and the projectile motion, with and without air resistance.

\subsubsection{Building Complexity: Planetary Orbits}
The third week starts with introducing students to Newton's magnum opus, the law of gravity presented in his {\em Treatise of the System of the World} \cite{Newton1728}. This law states that all objects with mass experience an attractive force toward other masses. Students learn that the acceleration due to gravity is not a constant but varies with the masses of the interacting bodies and the distance between them. They explore concepts such as Newton's cannonball, orbital velocity, and escape velocity.

With these new insights, students expand their projectile motion program to calculate the acceleration due to gravity at each time step of the simulation, thus enhancing their simulations with more sophisticated dynamics.
They analyze the effect of varying initial velocities and heights when launching an object at an angle with initial velocity, neglecting air resistance and the Earth's rotation. They also learn elements of numerical analysis, experiment with the Euler-Cromer algorithm for discretizing Newton's equations of motion, and explore how varying the time steps affects stability and numerical error.

By the end of the third week, the students' portfolios include a self-contained Jupyter notebook for modeling 2D motion above Earth, with varying gravitational acceleration, for various initial speeds and heights. They also work on a VPython program, to launch a rocket into space, between Earth and Venus. 

The last week is dedicated entirely to developing planetary orbits. Students begin by calculating the initial velocity for each planet using Newton's law of gravity, assuming circular orbits around the sun for simplicity. They are introduced to Kepler's laws of planetary motion to understand why planets move faster when closer to the sun and slower when further away. They calculate the orbital period for each planet using Kepler's third law, which states that the square of the orbital period of a planet is directly proportional to the cube of the semi-major axis of its orbit. Using this information, they determine how many time steps per orbit are needed to animate the motion of each planet with acceptable accuracy, learning to mitigate between the speed and stability of the simulation.

Students start their program with a single planet orbiting the sun, positioned at the origin of the coordinate system and kept fixed for simplicity. They then add the other three inner (or terrestrial planets), one at a time. More ambitious students may continue by adding the outer planets, despite the challenges posed by the large differences in distances and timescales.

Next, the students simulate the Earth--Moon system to observe how both bodies revolve around their common center of mass. They explore the effects of small deviations in their initial velocities and the numerical errors that accumulate over time, which can lead to elliptical, chaotic, or unstable orbits. They can improve their program by changing the numerical iterative algorithm and the time step to obtain a robust simulation.

With this knowledge gained, the ultimate challenge is to include the effects of mutual gravitational attraction between planets in their planetary system and study the behavior of the orbits over a relatively long simulation time, by simulating the Earth--Venus--Sun motion. 

By the end of the course, students have a deep understanding of motion and have developed a comprehensive number of projects, from the simple bouncing ball to the sophisticated dynamics involved in celestial mechanics.
Their portfolio contains four new VPython programs, from the Earth-Sun orbit and the planetary system, to the Earth--Moon and Earth--Venus--Sun motions.


\subsubsection{Revised Curriculum for a 15-Week Course}

The four-week, 4-hour/day curriculum presented above can be easily adapted into a fifteen-week, 3-hour/week semester course targeted towards freshman and sophomore students with no previous background in physics or computer programming.

In the first week students are introduced to the course, including installation and setup of necessary software. They begin with basic Python syntax, variables, and arithmetic operations to build a strong foundation. 
The next two weeks they learn the fundamentals of Python programming through Module 1 (01A, 01B, 01C). This module ensures that students grasp essential programming concepts and can apply them effectively.

Weeks four to five are dedicated to the physics of motion in one-dimension. Students apply the programming skills they have acquired and build their first simulation of a bouncing ball in VPython. 

Between weeks six and seven, students learn about scientific computation and code development. They begin by implementing two-dimensional motion with constant velocity and then progress to incorporating constant acceleration. 
Additionally, students complete the exercises in Module 03, which introduces iterative numerical methods for updating variables, and explore the impact of numerical methods like the Euler and Verlet algorithms on simulation accuracy, enhancing their understanding of computational techniques and their applications.

Weeks eight to nine continue with the exploration of two-dimensional motion. Students expand their bouncing ball simulation by implementing the projectile motion and ad the drag force to study the effect of air resistance. 

In weeks ten to eleven, students learn about Newton's law of gravity and the concept of variable acceleration. They apply these principles by building a simulation of the Newton's cannonball, and come to understanding the difference between orbital and escape velocities.
Then they analyze trajectories with different velocities and heights, incorporating elements of numerical analysis and stability conditions to ensure robust simulations and refine their projectile simulations to launch a rocket into space, between two planets.

Weeks twelve to fourteen are devoted entirely to building planetary orbits. Students start with a planet in simple circular orbit around the sun and gradually incorporate more planets at varying distances and velocities, increasing the complexity of their simulation. In the final week, students undertake a comprehensive project where they incorporate mutual gravitational attraction and study long-term orbital behavior. This project allows them to apply all the concepts and skills they have learned throughout the course, resulting in a comprehensive portfolio that demonstrates their proficiency in both computational programming and physics simulations.

This adapted curriculum ensures that the students progress at a manageable pace, building their skills incrementally while engaging in hands-on, interactive learning experiences. By the end of the semester, students will have a deep understanding of motion and the impact of numerical methods on simulations, culminating in a comprehensive portfolio that ranges from simple bouncing ball simulations to complex celestial mechanics.

The assessment is project-based and consists of six different sets of programs submitted at the end of each module, plus a final project. At the end of week 3, students turn in the homework for Module 1. By the end of week 5, they accomplish their first simulation of a bouncing ball in VPython. At the end of week 7, students are graded for Module 3 homework. By the end of week 9, they complete the 2D simulation of a projectile, both without and with air resistance. By week 11, students have developed a program for launching an object into space. By week 14, they have a set of programs animating planetary orbits. Instead of a comprehensive exam, students submit their final project, which may involve modeling a realistic scenario of launching a rocket with changing mass or including the effects of mutual gravitational attraction between planets.

\subsection{\label{sec:scicomp}Scientific Computing with Python}
\subsubsection{Setting the stage}
The narrative then advances to \quotes{Scientific Computing with Python} where sophomores and juniors in STEM are introduced to a variety of computational topics. The students form a heterogeneous group with different competencies, experiences, and interests. This course assumes only a background in vector calculus and linear algebra, and no more than a beginner’s experience in programming. 

The goal is not for students to become fluent in a computational language but to introduce them to computational modeling in science and teach basic programming structures and numerical algorithms. The aim is for students to understand fundamental numerical algorithms with applications to various real-world problems, and how to assess the uncertainty of their results.

Students interact with computational models in a scaffolded way, in a learning environment based on the \quotes{Two Bites at Every Apple} approach.
Students revisit and reinforce key concepts to ensure understanding and deepen learning, preparing them for more advanced challenges. For instance, elements from a planetary system program can be reused in a molecular dynamics calculation.

Students learn how to solve ordinary and partial differential equations encountered in biology, physics, chemistry, and engineering with computer programming. 
They use the powerful NumPy and SciPy Python libraries to perform numerical differentiation and integration, linear algebra and eigenvalue problems, Gaussian elimination, and iterative methods. The essential algorithms taught can solve a variety of problems and can be reused in other courses. Students understand how numerical algorithms solve mathematical problems, derive, verify, and implement these algorithms to achieve reproducible scientific outcomes, recognize potential issues, and learn how to think algorithmically to gain deeper insights into scientific problems.

Students work in small groups, relying on each other to develop an understanding of course material and computational elements.
More experienced students tutor others on different aspects of computational modeling, while less computationally-prepared students gain confidence on writing programs. They learn to run code, manage file structures and scripts, execute programs, collect data, test against exact solutions, and assess performance.

Through this process, students learn to develop mathematical and computational models of physical systems, design, implement, and validate functioning code, understand the role of numerical analysis in formulating and interpreting codes, use visualization and physical analysis to test hypotheses, and connect theory, experiment, and simulation. Additionally, students enhance their collaborative skills and develop a more profound appreciation for the interdisciplinary nature of computational science. This comprehensive approach prepares them for future academic and professional endeavors, where computational proficiency is increasingly essential.

\subsubsection{Course Development}
The \quotes{Scientific Programming with Python} course runs for one semester, using {\em Learning Scientific Programming with Python} by Christian Hill \cite{Hill2020} as the textbook, supplemented by its associated website \href{https://scipython.com/}{scipython.com}.
This course aims to equip students with basic programming skills in Python, focusing on powerful Python libraries: NumPy, SciPy, Matplotlib, and Pandas, to solve real-life problems in science and engineering. 

Given the mixed student population and the extensive content of the book, the course begins with Chapter 5, covering IPython and Jupyter Notebooks.
The first two weeks are dedicated to an introduction to IPython and Jupyter notebooks. Students learn to set up their environment, navigate basic commands, and create and run notebooks. This foundational knowledge prepares them for the subsequent modules. 

The rest of the curriculum is divided into four sections, each spanning approximately three weeks, where students are introduced to specific Python libraries and develop their skills through practical projects.

The first section, covering weeks 3 to 5, introduces the powerful NumPy library. The instruction focuses on array operations, file input/output, and on mathematical and statistical functions. During this period, students engage in numerical computations and perform array manipulations, building a solid foundation in handling large datasets and performing complex calculations efficiently.

The course then progresses to Matplotlib, where students start with basic plotting techniques and gradually move to more advanced data visualization, including 3D plotting and animations. Through hands-on exercises during weeks 6 to 8, students learn to create and customize plots, effectively visualizing and animating scientific data.

In the third section of the course, allocated for weeks 9 to 11, students explore the SciPy library, learning to solve differential equations and simulate real-world scientific problems. They engage with practical examples, such as modeling physical phenomena and performing numerical integration, enhancing their problem-solving skills with computational methods.

The final section, running from weeks 12 to 14, focuses on scientific data analysis using Pandas. Students learn to manipulate and analyze real datasets, gaining proficiency in data cleaning, transformation, and exploratory data analysis. They work on projects involving real-world data, preparing them for practical applications in scientific research and industry.

This curriculum is designed to be flexible. For instance, if students need to brush up on their Python programming skills, the last module covering the Pandas library can be omitted. The first three modules are then pushed back to accommodate a module on Python fundamentals during weeks 3 to 5. Additionally, group tutoring sessions can be offered at students' request. This ensures that all students, regardless of their initial proficiency, can keep pace with the course.

Throughout the course, students work on a final project, selecting and formulating a problem to solve using the concepts learned. This project involves developing a Python program that incorporates elements from NumPy, Matplotlib, SciPy, and Pandas libraries, culminating in a comprehensive demonstration of their skills. Students dedicate the last week to finalizing the project and presenting their results in front of the class.

\subsubsection{Scaffolded Approach}
The course employs a scaffolded approach, gradually building complexity and reinforcing concepts through the \quotes{Two Bites at Every Apple} method. This involves presenting an activity that addresses multiple audiences from different perspectives at the same time, revisiting key concepts in multiple contexts to deepen understanding, and is powerful for the mixed audience of students this course targets. Students use programming to solve simple problems such as predicting the outcome of the motion of an object under certain forces, visualization and data manipulation.

The course instruction is designed in workshop-style live programming environment, in which the instructor writes and executes code in real-time during the class, to teach interactively the programming concepts introduced and to engage students in real-time problem-solving. This method is an effective way to teach programming concepts interactively because it allows students to witness the entire process of problem-solving, writing, testing, and debugging the code, and it provides immediate feedback. By observing the thought process and techniques of the instructor, students can gain a deeper understanding of coding practices, learn to anticipate and resolve errors, and develop better problem-solving skills.

Students also learn important skills in data analysis, including reading, fitting, and manipulating data, using libraries such as NumPy, SciPy, and Pandas, through interactive projects such as a simple 2D diffusion model, linear and non-linear fitting of 2D data in SciPy, building heat maps, and a simple neural network for classification.

The assessment is based on weekly computer programs solving homework assignments from the textbook and the final project, which takes the place of the comprehensive exam. Students start working on this project early in the semester and can choose to extend the examples presented in class based on their interests. The topics for the final projects have broad scientific applications, from physics and engineering to chemistry and biology, depending on students' main field of study and interests. 

One example is the visualization and simulation of a particle moving under the effect of gravitational or electromagnetic forces. 
An extension of this project and a beautiful application of the Matplotlib library is the visualization of the Sun-Earth-Venus \quotes{rose} pattern by plotting Venus's motion in the Earth-Sun reference system.

Another project could entail the visualization of a vector field, with applications to the electric field of a capacitor and the dipolar magnetic field. This project can be continued with the gyro-motion of a particle in a magnetic field and extended to the motion of a particle in electric and magnetic fields and drift velocity.

Students then continue to the animation of 2D elastic collision of particles in a container, progressing to Brownian motion.
Another example involves calculating and plotting the trajectory of a spherical projectile, then including air resistance, and continuing with launching a satellite into orbit. 

These projects allow students to apply their computational skills to real-world problems, enhancing their understanding and preparing them for advanced scientific and engineering challenges.
By the end of the semester, students will have a strong foundation in scientific programming with Python, capable of applying their skills to a wide range of scientific and engineering problems, thanks to the scaffolded learning approach and repeated reinforcement of key concepts.

\subsection{\label{sec:compphys}Computational Physics}
\subsubsection{Setting the stage}
The culmination of this educational progression is \quotes{Computational Physics,} a course designed for advanced undergraduate and graduate students, primarily targeting math and physics majors. 
The student audience ranges from those with basic programming skills to those with significant experience. This course bridges a gap in the curriculum by teaching students how to apply computer programming to extend our scientific understanding of the world.
It introduces the computer as a tool for solving problems in physics, from the evolution of galaxies and propagation of gravitational waves to the complexities of the quantum world.

The main objective of this course is to familiarize students with advanced computational methods for solving data-intensive physics problems using symbolic and numerical programming. It emphasizes practical numerical, symbolic, and data-driven computational techniques used to predict the behavior of complex physical systems. 

The course focuses on advanced numerical algorithms and their applications in solving complex physics problems that are otherwise unsolvable without computational methods. 
Students develop the ability to solve real-world problems by deriving computational models from basic physics principles, constructing dimensionless or scaled models to simplify input data, and interpreting model assumptions to enhance understanding and predictions.
Preparing problems to be solved numerically is a critical step in readying students to solve computational problems in scientific settings. 
Students are introduced to computational thinking, that follows four stages: (1) breaking down complex problems into simple steps, (2) recognizing patterns and trends, (3) abstracting with functions or objects, and (4) developing algorithms \cite{Aiken2013}. They engage in reformulating scientific problems mathematically, selecting and implementing algorithms, visualizing and post-processing results, performing numerical analysis, assessing errors and stability, and using feedback to refine their approach.

The course covers advanced topics related to computational physics and immerses students in actual research, training them to develop more sophisticated programs and work on larger projects. 
The approach used is \quotes{Win-day-one} to ensure immediate engagement and practical experience from the first week. This method exposes students to high-level computational work right away, boosting their confidence and increasing their computational skills.
Instead of solving predefined computational problems, students are provided with snippets of \quotes{user-developed} code they might encounter in actual research. They are asked to extend these snippets to new situations, helping them understand what the program is doing and how it does it. This experience mirrors the common scientific and engineering practice of receiving and extending code from colleagues.

A \quotes{Win-day-one} exercise first provides learners with an answer, then methodically breaks down the code to explain each step. 
Initially, many details are abstracted away in functions and objects, but these are thoroughly explained in subsequent iterations involving multiple stages. 
Students are initially introduced to the complete workflow from start to finish, and subsequently they grasp in detail each component of the implementation in detail.
For example, in a numerical simulation of a colliding binary star system, using a finite difference approach, students design a mesh, form differential operators, set boundary conditions, generate a right-hand side, and solve the system. 

Students learn to work in a scientific computing environment, analyze physical problems, select appropriate numerical algorithms, and implement them using Python. They build algorithms to model and predict complex physical phenomena, apply data-driven computational approaches to analyze and visualize patterns, and demonstrate data analytical skills to identify signals from noise. Additionally, they learn to communicate data science methods and results effectively.
By the end of the course, students will have a deep understanding of computational physics, capable of applying their skills to a wide range of scientific and engineering problems.

\subsubsection{Course Details}

The \quotes{Computational Physics} course runs for one semester, offering three credit hours and targeting upper undergraduate and graduate students in physics or mathematics. Basic programming experience in Python is assumed, and college-level math and physics beyond the first year are required. To bring students up to speed in Python programming, the first three weeks utilize the book {\em A Student’s Guide to Python for Physical Modeling} by Jesse M. Kinder and Philip Nelson \cite{Kinder2021}. This guide provides a solid foundation in Python, specifically tailored for physical modeling, making it an excellent resource for students needing to strengthen their programming skills before diving into the course content.
In the spirit of the \quotes{Win-day-one} approach, in each of these first three weeks students jump directly into another \quotes{Computer Lab} chapter in the textbook, and turn to material from previous chapters if necessary as the coding process unfolds.

The main textbook is {\em Computational Modeling and Visualization of Physical Systems with Python} by Jay Wang \cite{Wang2016}, a  problem-oriented book that combines theoretical problems in physics with mathematical modeling and numerical techniques, to guide students to accurate computational solutions. 
The curriculum is restructured into five modules, each spanning about two weeks, covering topics from numerical methods for differential equations to boundary problems and statistical methods.

Instruction begins with basic numerical methods for Ordinary Differential Equations (ODEs), starting with simple Euler's method and progressing to the Leapfrog Method for enhanced stability. It then introduces Numerov’s Method, commonly used in orbital dynamics and quantum simulations, which handles second-order differential equations with high accuracy. During this phase, students simulate projectile motion with linear and quadratic air resistance, incorporating spin effects, and explore orbital mechanics through simulations of three-body motion influenced by gravitational forces.

The next module introduces advanced techniques for solving Partial Differential Equations (PDEs), including implicit methods, suitable for stiff equations that are challenging to solve numerically, the Finite Difference Method (FDM), and the Collocation Method, which employs radial basis functions for solving PDEs in irregular domains. Students engage in projects on mechanical wave propagation, using FDM to simulate waves on strings and membranes, including energy loss, and model the motion of charged particles in electric and magnetic fields in two and three dimensions.

Module three addresses boundary value and eigenvalue problems, introducing the Shooting Method—useful for converting boundary value problems into initial value problems—and Relaxation Methods, applicable in multi-dimensional boundary value problems. This period allows students to review and enhance their previous projects, work on method validation, and evaluate the stability and accuracy of the introduced numerical methods against analytical solutions.

In the fourth module, students learn two powerful statistical methods covering both stochastic and deterministic approaches to solving complex physical systems. They first study the Monte Carlo Method for probabilistic simulations to explore the statistical distribution and behaviors of random processes where exact solutions are analytically intractable. They then proceed with Molecular Dynamics Simulation, suitable for observing the behavior of large systems over time and under various conditions. Projects in this module include simulations of radioactive decay, random walks, Brownian motion in 1D and 2D, and the thermodynamics of interacting systems, such as achieving thermal equilibrium in Einstein solids.

The final module is dedicated to specialized numerical techniques, such as meshfree methods, which do not require a grid and use nodes randomly distributed in space, suitable for complex geometries and deformations where traditional mesh-based methods are inadequate. It also covers the Norm-preserving Method, specifically tailored for quantum mechanics simulations to ensure the conservation of probability. Applications in this module include structural analysis and fluid-structure interactions in molecular systems, together with quantum dynamics simulations of time-dependent quantum systems and quantum waves in two dimensions.

The course culminates in the last two weeks with students working on a final project of their choice, such as simulations of celestial few-body problems, molecular dynamics, linear and nonlinear oscillations, chaos and nonlinear dynamics, or quantum mechanics. It is common for students to apply the knowledge gained in this course to their comprehensive capstone research projects or master's theses, required for graduation, which can also serve as their final project for this course.

The course is structured around concise lectures, one-on-one interactions, and extensive individual work, both in and out of class. While essential physics concepts are introduced as needed, lectures do not cover all necessary information; students are expected to independently research and gather resources to complete their projects. They may make us of freely available online code, but must fully document its use.

Assessment is primarily project-based. Students are required to submit their codes and a short report in the form of a Jupyter notebook at the end of each module, detailing their findings and the methods employed. The culmination of the course is a final project, which serves as an alternative to the traditional comprehensive exam. This project is presented in class, allowing students to demonstrate their cumulative knowledge and skills applied in a practical context.

To provide a firsthand perspective on the impact of this course, below is a testimony from one of the students who found value in the applications of computational physics. 
\paragraph*{Student testimony:}
\quotes{My fascination with computational physics stems from its integration of various physics branches to tackle unsolved problems. This field allows me to explore complex, often extreme physical systems and contribute to our understanding of phenomena still mysterious to humanity. The challenge of exploring uncharted scientific territory is what excites me the most about computational physics.}


\section{\label{sec:projects}Students Accomplishments}

And now let's take a look at some of the accomplishments of various students that tool these courses. Below is a showcase of representative visualizations, taken mostly from the programs developed by students in the high-school/freshman \quotes{Physics Playground in Python} and the sophomore/junior \quotes{Scientific Programming with Python} courses.  

These visualizations highlight the practical application of computational techniques in physics, showcasing the students' ability to model, simulate, and visualize scientific problems. They reflect the skills acquired through hands-on projects and the scaffolded learning approach, demonstrating how students can effectively build computational models of various physical phenomena, from simple to complex.

\subsection{Physics Playground in Python Projects}
\begin{figure}[H]
\includegraphics[width=0.23\textwidth]{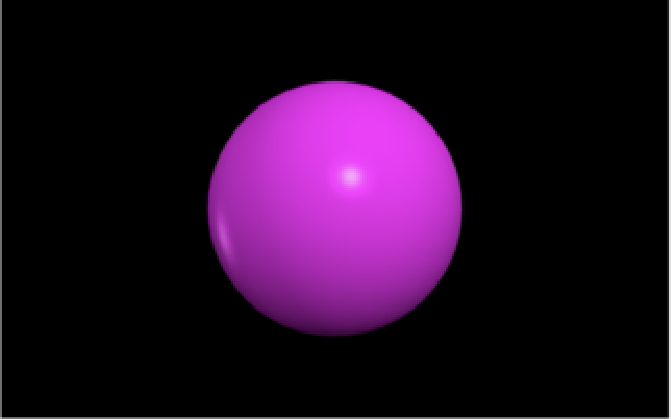}%
\quad
\includegraphics[width=0.23\textwidth]{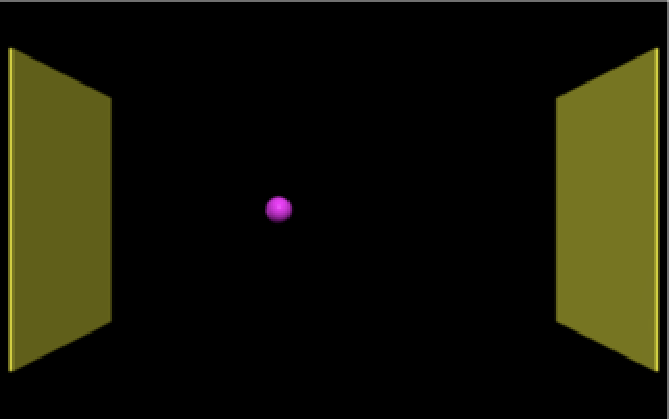}%
\caption{\label{fig:Figure1} Simulation of a ball moving (a) with constant velocity in one dimension and (b) reflected off two walls.}
\end{figure}

In Figure \ref{fig:Figure1} we see the first steps in simulating one dimensional motion with constant velocity, from a moving ball to a bouncing ball reflected off two walls. This was accomplished at the end of the first week/module in the \quotes{Physics Playground in Python} class. 

\begin{figure}[H]
\includegraphics[width=0.23\textwidth]{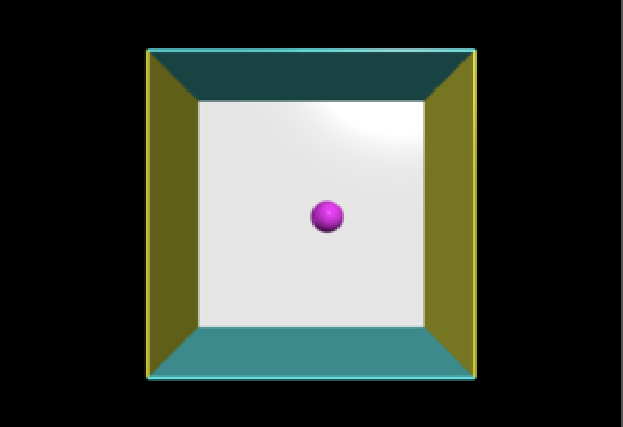}%
\quad
\includegraphics[width=0.225\textwidth]{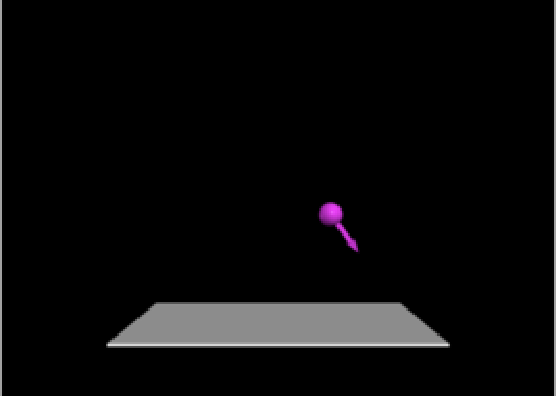}%
\caption{\label{fig:Figure2} (a) Simulation of a ball moving with 3D constant velocity and bouncing off inside a box.
(b) Projectile motion in 2D, with a dynamical arrow for the velocity.}
\end{figure}

Figure \ref{fig:Figure2} on the left, depicts the simulation of a ball moving with constant velocity in a tridimensional space and bouncing off the six walls of a box, with the front wall transparent.
On the right is a visualization of the projectile motion in 2D, with constant gravitational acceleration, including an arrow for the velocity that changes size and direction during motion.

\begin{figure}[H]
\includegraphics[width=0.23\textwidth]{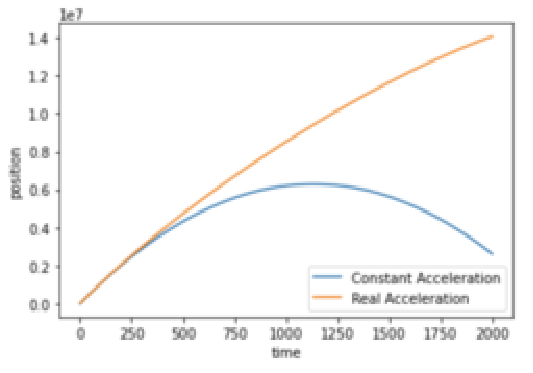}%
\quad
\includegraphics[width=0.225\textwidth]{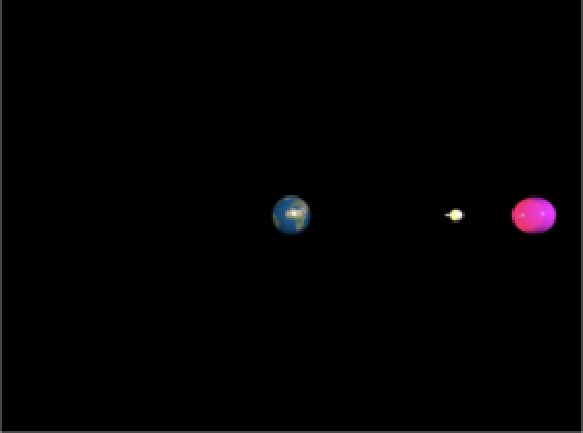}%
\caption{\label{fig:Figure3} (a) Simulation of a projectile with variable acceleration.
(b) Launch of a Satellite between Earth and Venus.}
\end{figure}

The simulation progresses in Figure \ref{fig:Figure3} to a projectile motion with gravitational acceleration depending on the height, shown on the left, for various initial velocities.
On the right we see the simulation of a satellite launched between Earth and Venus with escape velocity, in the simplified assumption that Earth and Venus do not move, and the force of the satellite on the planets is negligible. This is a one-dimensional three-body system, which is not solvable analytically due to the complicated interactions between the three bodies. Students explore its chaotic behavior, noting that it is very sensitive to initial data.

\begin{figure}[H]
\includegraphics[width=0.225\textwidth]{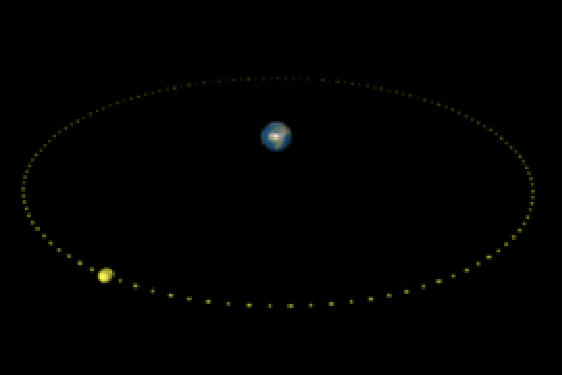}%
\quad
\includegraphics[width=0.23\textwidth]{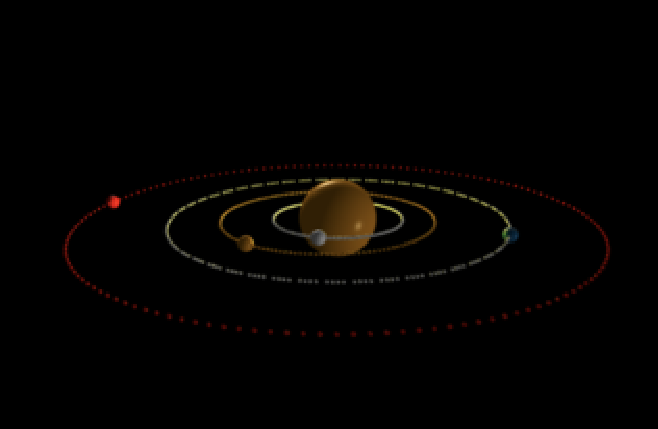}%
\caption{\label{fig:Figure4} (a) Simulation of Earth--Moon orbits around their common center of mass.
(b) Simulation of the four inner planets around the Sun}
\end{figure}

Figure \ref{fig:Figure4} contains two simulations of celestial dynamics: on the right the Earth--Moon orbit around their common center of mass, and on the left the orbits of the four inner terrestrial planets--Mercury, Venus, Earth and Mars--around the Sun, which is held fix, but Earth and Venus interact gravitationally.

The next figure, Figure \ref{fig:Figure5} is an ambitious visualization of the entire solar system, including the four outer giant planets, from Mercury to Neptune.
\begin{figure}[H]
\includegraphics[width=0.5\textwidth]{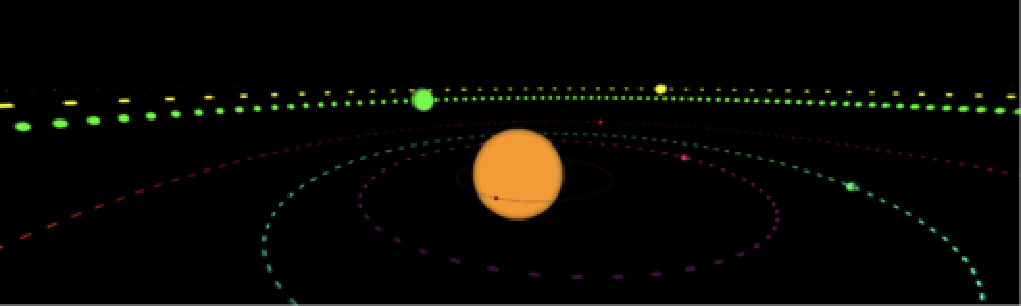}
\caption{\label{fig:Figure5}Simulation of the solar system.}
\end{figure}

\subsection{Scientific Computing with Python Projects}

The projects undertaken by students in the \quotes{Scientific Computing with Python} course vary by student interest. Below are a visualizations of a few representative students projects, using as start points programs available at \href{https://scipython.com/}{scipython.com}.

\begin{figure}[H]
\includegraphics[width=0.21\textwidth]{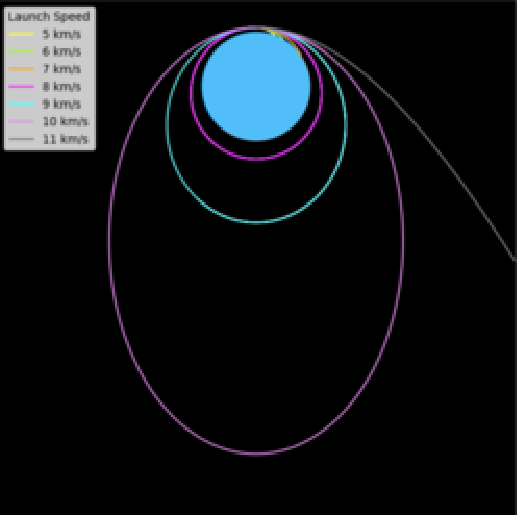}%
\quad
\includegraphics[width=0.23\textwidth]{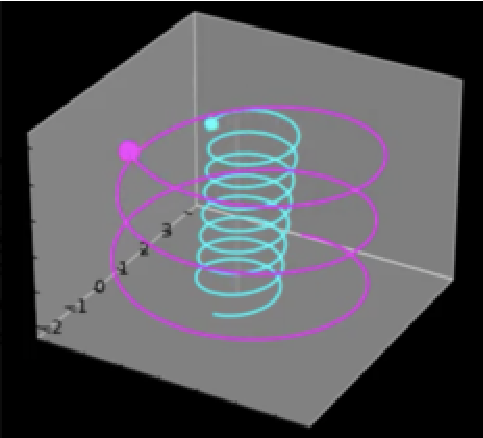}%
\caption{\label{fig:Figure6} (a) Simulation of a rocket launched above Earth with different speeds. 
(b) Simulation of an electron and a positive ion moving in perpendicular electric and magnetic fields.}
\end{figure}

Figure \ref{fig:Figure6} contains two examples in which students use the NumPy and SciPy libraries to calculate and plot trajectories of an object under the action of forces in two and three dimensions.
The figure on the left shows the motion in gravitational field of a rocket launched from a certain distance above the Earth's surface, for four different initial speeds, ignoring atmospheric drag and the Earth's own rotation.  
On the right, it depicts the motion of two particles of equal but opposite charges and different masses released with the same velocity in a perpendicular configuration of electric and magnetic fields, subjected to a Lorentz force.

\begin{figure}[H]
\includegraphics[width=0.25\textwidth]{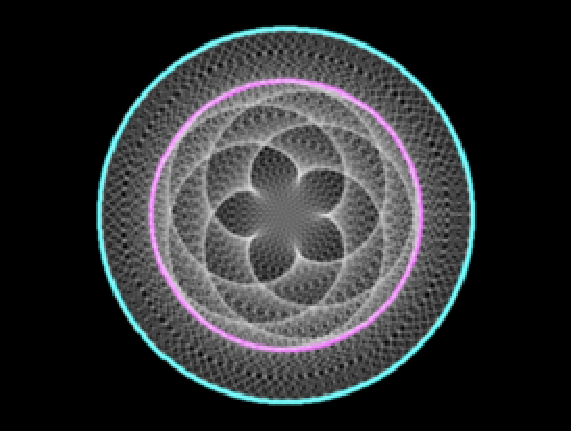}%
\quad
\includegraphics[width=0.19\textwidth]{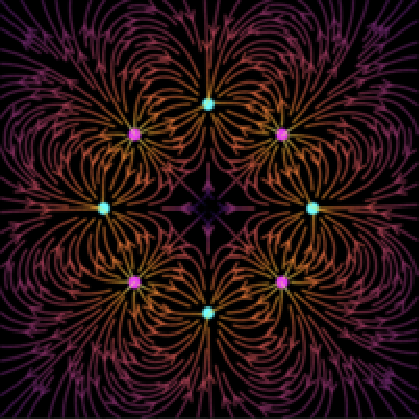}%
\caption{\label{fig:Figure7} (a) The Earth--Venus dance around the Sun.
(b) Diagram of the electric field vector of a quadrupole of charges.}  
\end{figure}

In Figure \ref{fig:Figure7} students use the Matplotlib library to create visually pleasing geometries of celestial mechanics, representing the "dance of planets," and to diagram electric and magnetic vector fields. On the left is a representation of the Earth-Venus dance, or the "rose" of Venus, obtained by joining the positions of Earth and Venus with a straight line as they orbit around the Sun. The right plot provides insight into the electric field vector configuration around a multipole arrangement of charges.

\subsection{Computational Physics}

\begin{figure}[H]
\includegraphics[width=0.5\textwidth]{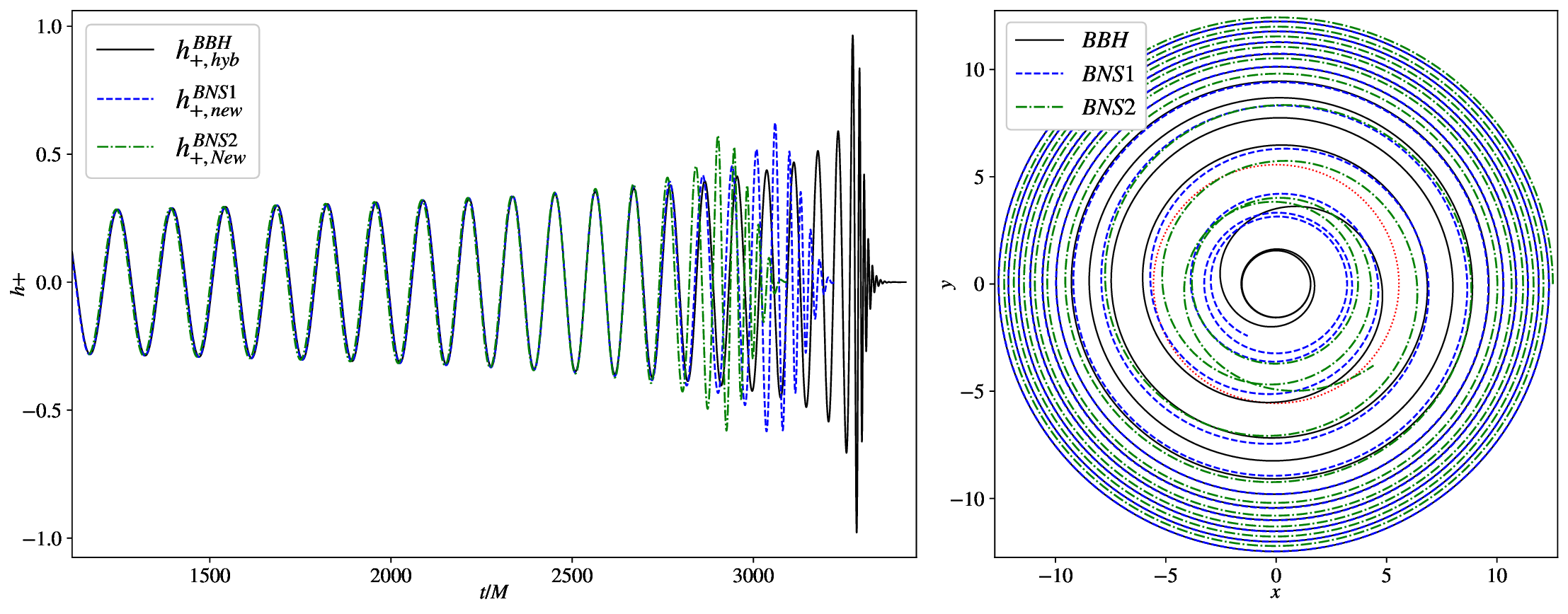}%
\caption{\label{fig:Figure8} (a) Gravitational wave strain for a binary black hole (solid black), and two binary neutron stars with different equations of state (dotted blue and dashdot cyan). 
(b) The separation between the stars in the binary up to merger. \cite{ODell:2023}}
\end{figure}

In the \quotes{Computational Physics} class, students focus less on visualization and more on obtaining accurate simulations of computationally challenging problems without analytically known solutions. Most of the work students do in this class is included in their final capstone or master thesis, with some resulting in conference presentations and publications, such as \cite{Buskirk2018, Buskirk:2022, ODell:2023}.
Figure \ref{fig:Figure8} is an example from a project where an undergraduate student researched post-Newtonian simulation of binary black holes and neutron star collisions, aiming to model the effect of matter on emitted gravitational waves.
It shows on the left the numerical calculation of the gravitational wave strain for a binary black hole (solid black), and two binary neutron stars with different equations of state (dotted blue and dashdot cyan) using post-Newtonian theory, and on the right the evolution of the separation between two orbiting neutron stars for the two equations of state, continued to the merger.

\section{\label{sec:concl}Conclusion}

In this chapter, we illustrated how computational methods can be integrated into the physics curriculum using a scaffolded approach across three distinct courses—each tailored to different levels of student expertise and offered in succession. The curriculum starts with an introductory level class for freshmen and high school seniors, relying on the \quotes{Hello World} teaching method. This is followed by an intermediate, sophomore-level class implemented in the \quotes{Two Bites at Every Apple} learning style, and culminates with a senior/graduate-level class using the \quotes{Win-day-one} approach.

The hidden goal of the above course sequence is to facilitate a change in students' attitudes towards computational physics through engagement and to attract students who might not have considered this path. The need for computational skills is dire in the workplace, and by showing our students that we care about their future careers, we aim to make a difference in the way students relate to the physics major.

We have detailed the steps of the implementation, from simple and engaging simulations to advanced projects that prepare students for real-world scientific research. We showcased some of the students' results, reflecting their journey from basic programming to complex problems in computational physics. This scaffolded approach has the potential to enhance students' understanding of how computational methods can be applied to physical principles and can form the basis for a minor or even a major or concentration in computational physics.

One drawback is that these classes suffer from low enrollment, being elective, and students take them either to complete an elective requirement or if required by their capstone mentor. This gap between the need for computational skills in the workplace and student enrollment could be alleviated if these courses become a required part of the STEM curricula.

This curriculum is flexible and can be adapted to either smaller classrooms or large-scale programs. Small schools can tailor these teaching approaches in computational physics to accommodate smaller settings, varying backgrounds, limited resources, and diverse student populations and skill levels. Large schools can incorporate hands-on, personalized learning approaches into larger courses by creating modules focused on computational physics within broader programs or using independent study models for advanced classes to offer research-oriented computational-physics experiences.

Another challenge is keeping the curriculum up-to-date with advances in computational physics. It is important to ensure that instructors keep up with this evolving field and remain proficient. As Generative Artificial Intelligence (AI), such as ChatGPT, becomes increasingly common, they will likely change how we think of computational physics. Currently, generative AI can produce code, and this will necessarily impact how students learn computational physics, potentially hindering their own learning by using LLMs to complete assignments. One option is to welcome change and start thinking about teaching students how to effectively use LLMs to solve computational problems and train the next generation of physicists with the tools and skills necessary for keeping pace with cutting-edge research.
Embracing these changes will ensure that students are well-equipped to contribute to scientific and technological innovations.

Looking forward, the integration of computational physics into education continues to hold great promise.
As computational methods evolve and become more sophisticated, their application in physics education will likely expand. 
This integration not only prepares students for modern scientific challenges but also enhances their problem-solving skills, making them more competitive in the job market. 

\begin{acknowledgments}
This work was funded by jointly a PICUP stipend to support computation-based curricular change, and by Marshall University, and in part by the National Science Foundation under Grant No. NSF PHY-1748958.
\end{acknowledgments}



\bibliography{references}
\end{document}